# Approximating the Lateral Distribution Function of Cherenkov Radiation as a Function of the Particle Type for Tunka-133 Array

**Zena Fadhel Khadhum[1], Hassan Abdullah Mahdi[2], A.A. Al-Rubaiee[2,*]**

[1]*College of science, Department of astronomy and space, Baghdad University, Baghdad, Iraq*
[2]*College of Science, Department of Physics, Al-Mustansiriyah University, Baghdad, Iraq*

*Email: dr.rubaiee@uomustansiriyah.edu.iq

**Abstract**
The main interest of the present work is in analyzing the lateral distribution function (LDF) of Cherenkov radiation from particles that produced in Extensive Air Showers (EAS). The simulation of LDF of Cherenkov radiation is fulfilled by utilizing the CORSIKA program at $3 \cdot 10^{15}$ eV of the primary energy around the knee region for many primaries for vertical showers for Tunka-133 array conditions. Depending on the numerical simulation results of Cherenkov light LDF, sets of parameterized polynomial functions are resetted for several particles as a function of primary particle type. The comparison between the approximated LDF of Cherenkov radiation with the LDF which has been simulated using CORSIKA program for Tunka-133 array is verified for several primary particles for vertical EAS cascade.

## 1. Introduction

Primary cosmic rays (PCRs) of energy spectrum and mass composition in EAS have been studied around the knee region. This approach has a main significant for getting information about PCR acceleration mechanisms and origin [1, 2]. Generally, LDF of Cherenkov radiation depends on the type and energy of the produced primary particle, observation level, distance from EAS core $R$, height of the first interaction and the direction of shower axis [3, 4]. Various studies on Cherenkov radiation in EAS have been performed by many authors. V. Prosin et al. have presented the results of studying CRs energy spectrum and mass composition over 3 seasons with Tunka-133 EAS array [5]. Furthermore, they have improved a method of EAS parameter reconstruction. G. Rastegarzadeh et al. have compared the Cherenkov light LDF results simulation for gamma and hadron showers with the experimented results of Tunka-like array in the energy range 100GeV-25TeV [6]. On the other side, I. De Mitri has studied CRs energy spectrum using ARGO-YBJ experiment in the energy range (TeVs$\rightarrow$ PeVs) [7].
In this paper, the LDF simulation of Cherenkov radiation was fulfilled utilizing CORSIKA code around the knee region for configurations of Tunka-133 array [8, 9] at the energy 3PeV for vertical showers and several primary particles, such as: (H, He, Li, Be, Ne, Na, Mg, Al, Sc, Ti, V and Cr). An approximating of Cherenkov light LDF has been obtained by depending on Breit-Wigner model [10, 11], by basing on numerical simulation method as a function of the particle's type.

## 2. Lateral Profile of Cherenkov Light

Cherenkov radiation can be emitted by ultrahigh energy particles (such as electrons, positrons and muons) of developing EAS, which gives information about the particle producing the cascade. Charged particle threshold energy, that excite the Cherenkov light in the atmosphere might be specified by the condition $\gamma = \gamma_{th}$ with $\gamma = E/mc^2$, where $\gamma$ is the particles Lorenz factor in the laboratory system. At the threshold energy, Lorenz factor ($\gamma_{th}$) can be given as [12]:

$$\gamma_{th} = \frac{1}{\sqrt{1-\beta^2}} = \frac{1}{\sqrt{1-(1/n(h))^2}} \tag{1}$$

The condition $\gamma > \gamma_{th}$ is equal to the condition $v > c/n$, i.e. $n\beta > 1$, where $\beta = v/c$, that means, the charged particle speed must exceed the speed of light $c$ with $n$ refractive index that depends on the height of shower ($h$) [12]:

$$n(h) = 1 + \zeta(h),$$

Where,

$$\zeta(h) = \zeta_0 \exp\left(-\frac{h}{h_0}\right), \tag{2}$$

When $\zeta_0 \approx 3 \cdot 10^{-4}$, $h_0$=7.5 km. The energy of electron in atmosphere is given as [13]:

$$E_{th} = mc^2 \gamma_{th} \tag{3}$$

By substituting equation (1) in equation (3) one can get:

$$E_{th} = \frac{mc^2}{\sqrt{1-(1/n(h))^2}} = mc^2 \frac{n(h)}{\sqrt{n(h)^2 - 1}} \tag{4}$$

The threshold energy of electron, which depends on the height ($h$) in atmosphere, can be given through eq. (2) and eq. (3) where:

$$E_{th}(h) = mc^2 \frac{1+\zeta}{\sqrt{2\zeta(1+\zeta/2)}} \approx \frac{mc^2}{\sqrt{2\zeta_0}} \exp(h/2h_0) \tag{5}$$

The Cherenkov light wil emit an electrons with energies greater than $E_{th}$. Thus at sea level $n = 1 + \zeta_o$ and $\gamma > \gamma_{th} = \frac{E_{th}}{mc^2} \approx 40.8$ [12].

For Cherenkov photons radiation by electrons, $E_{th}$ in the atmosphere at the highest $h$ approximately canbe defined from equation (5) at the height $h_o = 7.5\ km$.
Cherenkov radiation that initiated by particles with very high energies $\beta \approx 1$ slants under a small incident angle $\theta_r$, that is given by the relation [12]:

$$cos\theta_r = \frac{1}{\beta n} \approx 1 \tag{6}$$

From eq. (1) to eq. (6), one can find:

$$sin^2_{\theta_r} = 2\zeta_0 \exp\left(-\frac{h}{h_0}\right)\left[1 - \frac{E_{th}^2(h)}{E^2}\right], \tag{7}$$

At sea level, when $E >> E_{th}$ (h), one can find:

$$sin\theta \approx \theta_r \approx \frac{1}{\gamma_{th}} \approx \sqrt{2\zeta_0} \exp\left(-\frac{h}{2h_0}\right) = 2.45 \cdot 10^{-2} \quad (8)$$

At the wavelength interval $(\lambda_1, \lambda_2)$, the Cherenkov photons quantity can be found through the Tamm-Cherenkov relation [14], where:

$$\frac{dN_\gamma}{dx} = 2\pi\alpha sin^2_{\theta_r} \int_{\lambda_1}^{\lambda_2} \frac{d\lambda}{\lambda^2} = 2\pi\alpha \left(\frac{1}{\lambda_1} - \frac{1}{\lambda_2}\right) \zeta_0 \left(1 - \frac{E_{th}^2(h)}{E^2}\right) \exp\left(-\frac{h}{h_0}\right) \quad (9)$$

Where $\alpha = 1/137$. The rate of the Cherenkov photon is:

$$\frac{dN_\gamma}{dx} = \frac{dN_\gamma}{dx}\frac{dx}{dt} = x_0 \exp\left(\frac{h}{h_0}\right)\frac{dN_\gamma}{dx}, \quad (10)$$

Where
$$\frac{dx}{dt} = x_0 \exp\left(\frac{h}{h_0}\right), \quad (11)$$

Where $x_0 = \frac{t_0}{\rho_0} \approx 3.09 \cdot 10^4$, that is defined as the electrons distance at sea level; $t_0 = 37.1$ and $\rho_0 = 1.2 \cdot 10^{-3}$ [14].

The total number of photons $N_\gamma$ that radiated by electrons with neglecting the absorption process in atmosphere, can be defined as:

$$N_\gamma = 45 \cdot 10^{10} \frac{E_0}{10^{15} eV} \quad (12)$$

The LDF is a function that describes the lateral variation of Cherenkov flux with a distance from the shower axis, which is used extensively in event reconstruction, for getting an information about primary particle. Estimating of the age parameter and the core position are made also through utilizing the total number of photons in the cascade shower that is given as: $(E_0)$:

$$N_\gamma = 3.7 \cdot 10^3 \frac{E_o}{\beta_t}, \quad (13)$$

where $\beta_t = \beta_{ion} t_o$, which it is the critical energy. For electron, $\beta_{ion} = 2.2 \text{ Mev.} (g. \text{ cm}^{-2})^{-1}, t_o = 37 \text{ g. cm}^{-2}$ and $\beta_t = 81.4 \text{ MeV}$ [14].

In general case, the LDF of Cherenkov radiation is depended on the primary particle energy that produced in EAS. The photons number per detector area unit is given by [12]:

$$Q(E,R) = \frac{\Delta N_\gamma(E,R)}{\Delta S} \quad (14)$$

where $\Delta N_\gamma$ is number of photons, $\Delta S$ is the area of detector.

### 3. Results and Discussion
### 3.1 Modelling and Approximating of Cherenkov Light LDF

The LDF of Cherenkov light simulation is fulfilled using a detailed Monte Carlo program that is called CORSIKA code (Cosmic Ray Simulation for KAscade) [15], which studying the development and characteristics of EAS produced by energies up to $10^{20}$ eV. This simulation was fulfilled for conditions of Tunka-133 array around the knee energy region (3 PeV) for several primary particles. Two hadronic models were used through the simulation; QGSJET code [16] that was used for modeling the hadrons interactions with high energies and GHEISHA code [17] that was used for modeling the hadrons interactions for lower energies. For approximating the simulated LDF of Cherenkov radiation. The proposed four parameters function are used and given by [18, 19]:

$$Q(\eta, R) = \frac{C\sigma exp[\alpha - \xi]}{\beta[(R/\beta)^2 + (R-k)^2/\beta^2 + R\sigma^2/\beta]} \quad (15)$$

Where $\xi$ is a parameter given by:
$$\xi = R/\beta + (R - k)/\beta + (R/\beta)^2 + (R - k)^2/\beta^2 \quad (16)$$

where $C$ is the normalization constant [7]; $R$ is the distance from the shower axis; $\alpha, \beta, \sigma$ and $k$ are parameters of Cherenkov light LDF, which are approximated as a function of the primary particle, which is given by the relation:

$$Y(\eta) = c_o + c_1 log_{10}(\eta) + c_2 (log_{10}(\eta))^2 + c_3 (log_{10}(\eta))^3 \quad (17)$$

where $Y(\eta)$ is presented for the parameters: $\alpha$, log $\beta$, log $\sigma$ and log $k$ (see Table 1). Whereas, $c_0$, $c_1$, $c_2$, and $c_3$ are coefficients depended on the primary particle's type and presented in see Table 1.

**Table 1:** Coefficients $c_i$ which determine $\eta$ and depending on $\alpha, \beta, \sigma$ and $k$ parameters for vertical shower

| | $\theta=0°$ | | | | |
|---|---|---|---|---|---|
| Y | $c_0$ | $c_1$ | $c_2$ | $c_3$ | $x^2$ |
| $\alpha$ | $78.654.10^{-1}$ | $19.775.10^{-1}$ | $-73.510.10^{-2}$ | $83.960.10^{-3}$ | $37.3.10^{-4}$ |
| $\beta$ | $-49.106.10^{-2}$ | $11.263.10^{-1}$ | $-45.005.10^{-2}$ | $55.050.10^{-3}$ | $34.5.10^{-4}$ |
| $\sigma$ | $29.746.10^{-1}$ | $-45.592.10^{-1}$ | $18.961.10^{-1}$ | $-24.605.10^{-2}$ | $23.6.10^{-4}$ |
| $k$ | $-18.761.10^{-1}$ | $-12.180.10^{-3}$ | $28.143.10^{-2}$ | $-58.120.10^{-3}$ | $57.0.10^{-5}$ |

The approximated Chrenkov light LDF in Figures 4, 5 and 6 for vertical cascades is slightly differs from that simulated for conditions of Tunka-133 array. The difference between the calculated and simulated LDF of Cherenkov radiation is specified by minimization of the formula:

$$\Delta = \sum_i \left[\frac{Q_{par}(\eta, R)}{Q_{COR}(R)} - 1\right]^2 \rightarrow min, \quad (18)$$

Where, $Q_{par}$ is the parameterized Cherenkov light LDF by eq. (15) and is the simulated Cherenkov light LDF by CORSIKA code. In Table 2 was demonstrated the results of minimization for each primary particle by using eq. (18).

Figures (4-6) demonstrate the parameterization of Cherenkov light LDF using Eqs. (15)-(17) in comparison with that simulated with CORSIKA programe for vertical showers and several primary particles at the energy 3 PeV.

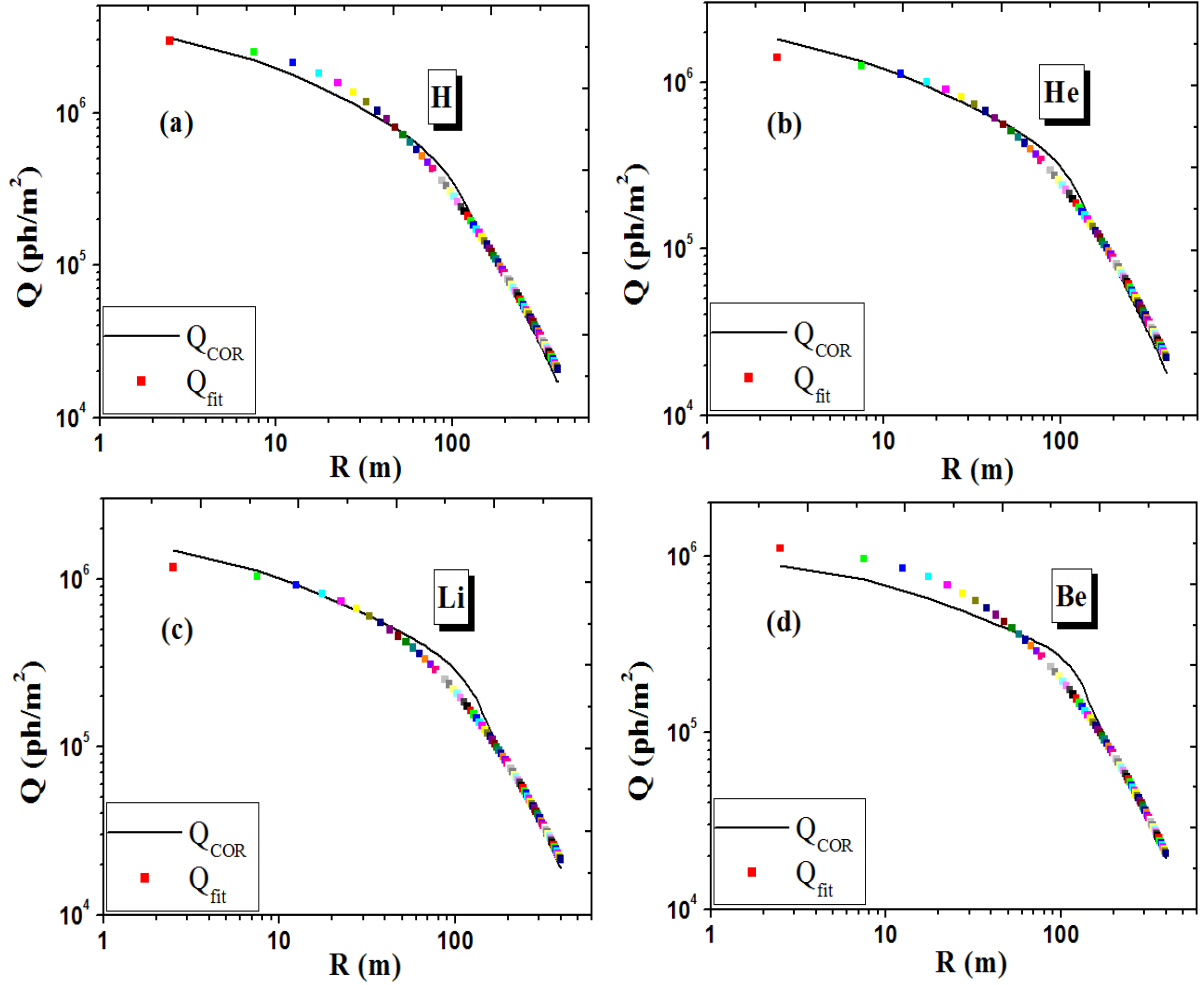

**Fig. 4** Cherenkov light LDF simulation by CORSIKA program at 3 PeV of primary energy with θ=0° (solid curves) in comaparision with that parameterized using Eq. (15) (symbols) for: (a) H; (b) He; (c) Li and (d) Be. Where R(m) is the distance from the showers axis in meter units.

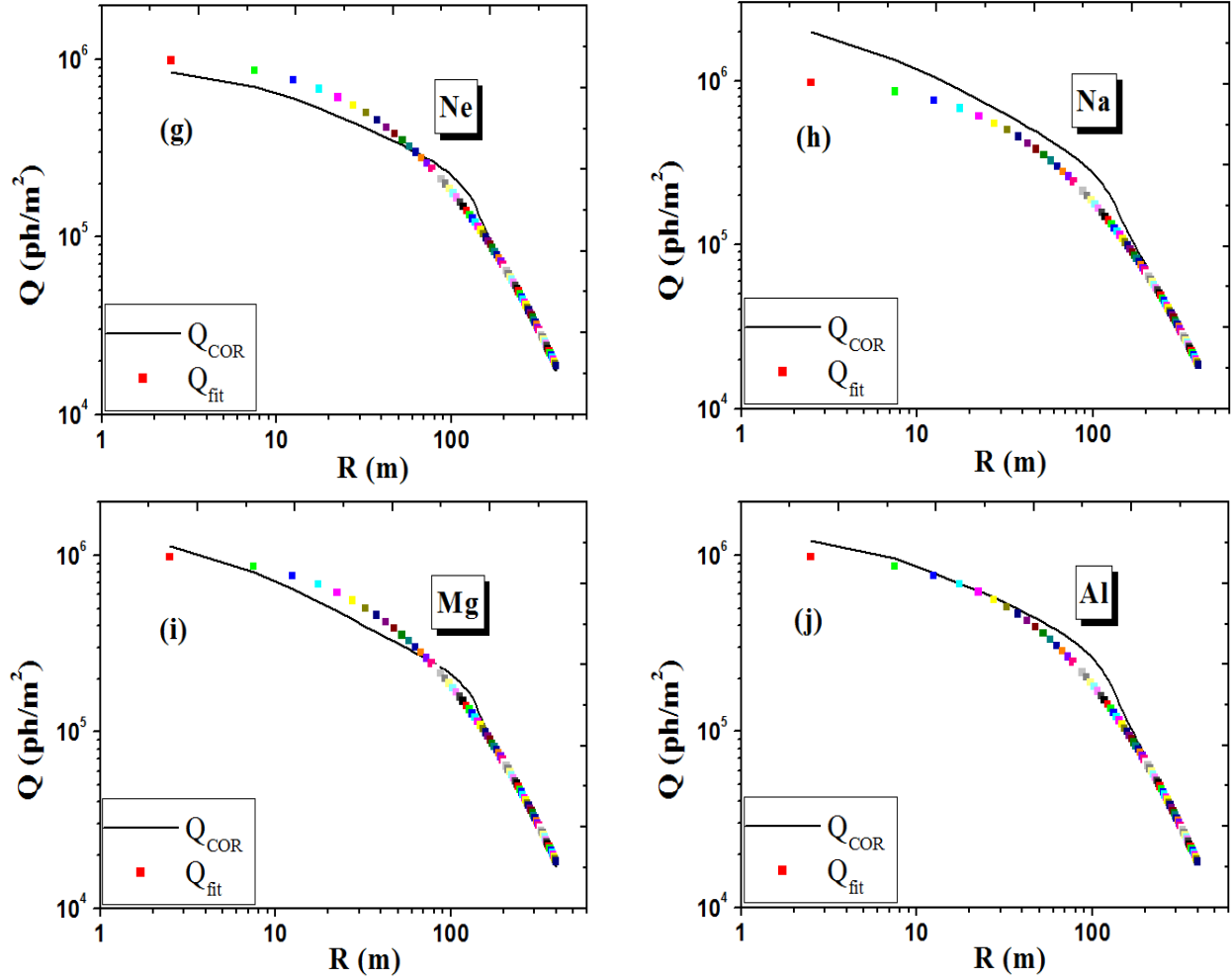

**Fig. 5** Cherenkov light LDF simulation by CORSIKA program at 3 PeV of primary energy with θ=0° (solid curves) in comaparision with that parameterized using Eq. (15) (symbols) for: (g) Ne; (h) Na; (i) Mg and (j) Al. Where R(m) is the distance from the showers axis in meter units.

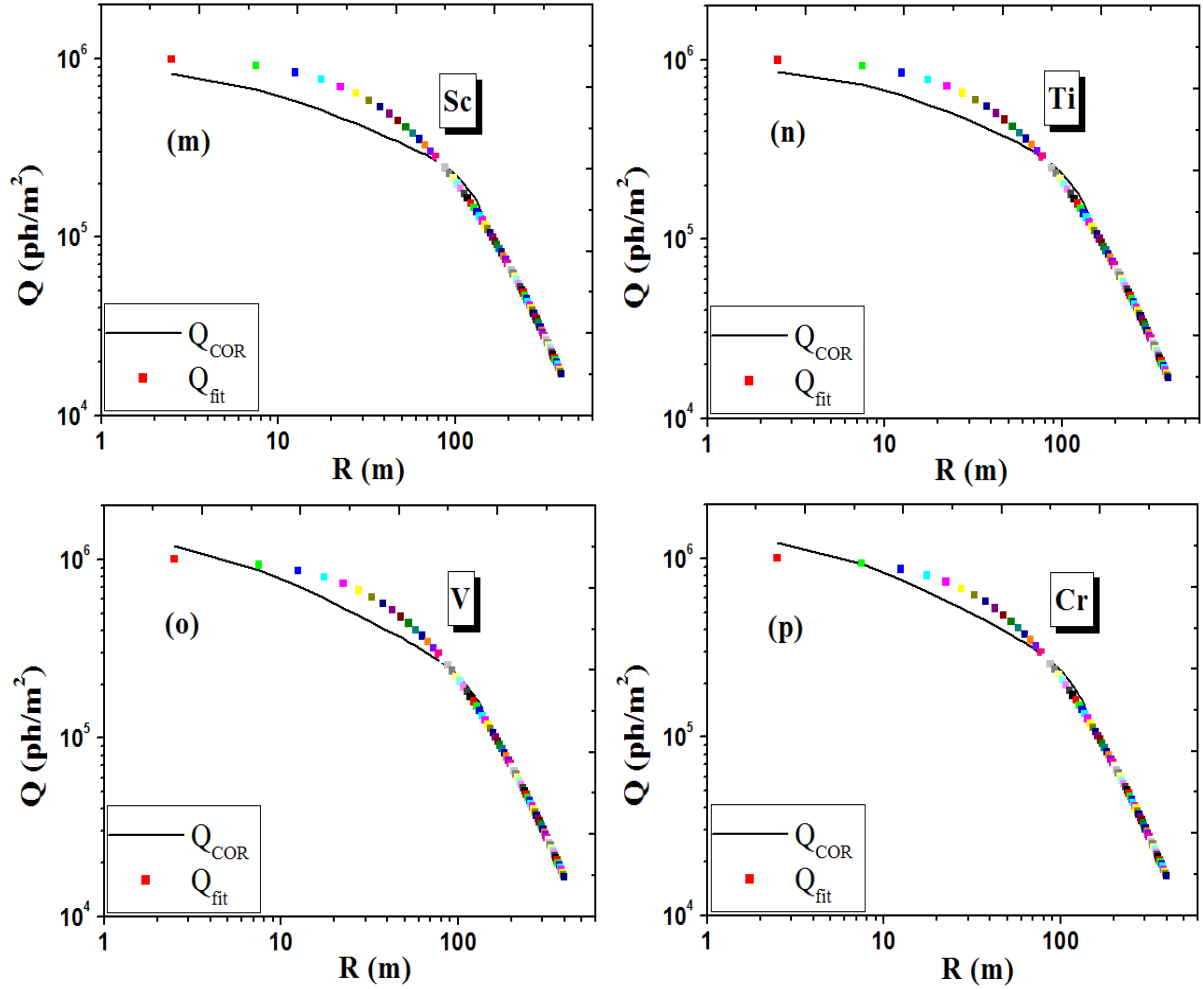

**Fig. 6** Cherenkov light LDF simulation by CORSIKA program at 3 PeV of primary energy with θ=0° (solid curves) in comaparision with that parameterized using Eq. (15) (symbols) for: (m) Sc; (n) Ti; (o) V and (p) Cr. Where R(m) is the distance from the showers axis in meter units.

**Table 2:** $\Delta_{min}$ results that estimated with Eq. (18) for different particles at 3 PeV primary energy.

| Element | Symbol | H | $\Delta_{min}$ |
|---|---|---|---|
| *Hydrogen* | **H** | **101** | $14.699 \cdot 10^{-4}$ |
| *Helium* | **He** | **402** | $22.873 \cdot 10^{-4}$ |
| *Lithium* | **Li** | **703** | $85.901 \cdot 10^{-4}$ |
| *Beryllium* | **Be** | **905** | $12.755 \cdot 10^{-4}$ |
| *Neon* | **Ne** | **2010** | $22.071 \cdot 10^{-4}$ |
| *Sodium* | **Na** | **2310** | $71.613 \cdot 10^{-3}$ |
| *Magnesium* | **Mg** | **2412** | $21.778 \cdot 10^{-4}$ |
| *Aluminum* | **Al** | **2711** | $16.147 \cdot 10^{-3}$ |
| *Scandium* | **Sc** | **4517** | $28.923 \cdot 10^{-3}$ |
| *Titanium* | **Ti** | **4810** | $16.649 \cdot 10^{-3}$ |
| *Vanadium* | **V** | **5117** | $12.656 \cdot 10^{-3}$ |
| *Chromium* | **Cr** | **5224** | $35.013 \cdot 10^{-4}$ |

**Conclusion**

The lateral profile of Cherenkov radiation in EAS initiated by several primaries such as (H, He, Li, Be, Ne, Na, Mg, Al, Sc, Ti, V and Cr) has been simulated by CORSIKA program at the energy 3 PeV for conditions of Tunka-133 array. Basing on the simulated results, sets of parameterized functions were constructed by depending on Breit-Wigner model as a function of particle's type. The comparison between the approximated lateral distributions of Cherenkov radiation with that simulated using CORSIKA code for Tunka-133 array is verified for several primary particles. This comparison proved the ability of particle identification producing an extensive air showers and energy determination and mass composition around the knee region of the cosmic ray spectrum. This approach gives a significane of making a library of lateral distributions of Cherenkov radiation within a short time that could be utilized for real events analysis that have been detected in the experimental arrays and for energy spectrum and mass composition reconstructing for high energy cosmic rays.